# Ethics as a service: a pragmatic operationalisation of AI Ethics

Jessica Morley[1], Anat Elhalal[2], Francesca Garcia[2], Libby Kinsey[2], Jakob Mökander[1], Luciano Floridi[1,3]

[1.] Oxford Internet Institute, University of Oxford

[2.] Digital Catapult, London

[3.] Alan Turing Institute, London

## Abstract

As the range of potential uses for Artificial Intelligence (AI), in particular machine learning (ML), has increased, so has awareness of the associated ethical issues. This increased awareness has led to the realisation that existing legislation and regulation provides insufficient protection to individuals, groups, society, and the environment from AI harms. In response to this realisation, there has been a proliferation of principle-based ethics codes, guidelines and frameworks. However, it has become increasingly clear that a significant gap exists between the theory of AI ethics principles and the practical design of AI systems. In previous work , we analysed whether it is possible to close this gap between the 'what' and the 'how' of AI ethics through the use of tools and methods designed to help AI developers, engineers, and designers translate principles into practice. We concluded that this method of closure is currently ineffective as almost all existing translational tools and methods are either too flexible (and thus vulnerable to ethics washing) or too strict (unresponsive to context). This raised the question: if, even with technical guidance, AI ethics is challenging to embed in the process of algorithmic design, is the entire pro-ethical design endeavour rendered futile? And, if no, then how can AI ethics be made useful for AI practitioners? This is the question we seek to address here by exploring why principles and technical translational tools are still needed even if they are limited, and how these limitations can be potentially overcome by providing theoretical grounding of a concept that has been termed 'Ethics as a Service'

## 1. Introduction

As the range of potential uses for Artificial Intelligence (AI), in particular machine learning (ML), has increased, so has awareness of the ethical issues posed by the design, development, deployment and use of AI systems (henceforth collapsed into 'Design'). Issues such as privacy, fairness, accountability, accessibility, environmental sustainability, and transparency are now not just discussed in academic literature but also in mainstream media. This increased awareness has led to the realisation that existing 'hard' governance mechanisms (such as legislation and other regulatory frameworks, e.g. ISO requirements) alone provide insufficient protection to individuals, groups, society, and the environment. Similarly, these mechanisms alone do not sufficiently incentivise the Design of socially preferable and environmentally sustainable AI. In an attempt to overcome these limitations, governments, private sector organisations, and others have focused on the development of 'soft' governance mechanisms such as ethics codes, guidelines, frameworks, and policy strategies (Floridi, 2018; Schiff et al., 2020). The development of these largely principle-based documents has been an important and necessary phase in the evolution of AI governance (Mulgan, 2019; Raab, 2020). However, it has become increasingly clear that highly abstract principles provide little protection from potential harms related to AI when AI practitioners have no guidance on how to design and deploy algorithms within these ethical boundaries (Clarke, 2019; Orr & Davis, 2020). In other words, a significant gap exists between theory and practice within the AI ethics field (Ville et al., 2019). This is not unusual in ethics (consider for example the development of bioethics), where changes are sometimes theory-led, and can partly be explained by the relative 'newness' of the concept of AI ethical principles in the public policy domain[1]: less than 20% of all the AI ethics documents are more than four-years old (Jobin et al., 2019). However, it may also be a result of the desire by influential private-sector organisations to 'ethics wash' (Floridi, 2019b) in an attempt to keep the ethics of AI a self-regulated field and delay legislative intervention (Butcher & Beridze, 2019).

In previous work (Morley et al., 2019), we analysed whether it may be possible to start closing this gap between the 'what' and the 'how' of AI ethics by identifying the methods and tools already available to help AI developers, engineers, and designers (collectively 'practitioners' (Orr & Davis, 2020)) know not only what to do or not to do, but also how to do it, or avoid doing it, by adopting an ethical perspective (Alshammari & Simpson, 2017). We plotted the tools in a typology, matching them to ethical principles (beneficence, non-maleficence, autonomy, justice and explicability) and to stages in the algorithm development pipeline. Although we found that numerous tools and methodologies exist to help AI practitioners translate between the 'what' and the 'how' of AI ethics, we also found that the vast majority of these tools are severely limited in terms of usability. The development of these translational tools and methods may have been useful for enabling individual groups of researchers/companies to raise internal awareness of AI ethics and to examine different interpretations of ethical principles. However, this impact has not been sufficiently tested and the external validity of all the tools/methods identified remains questionable. There is, as of yet, little evidence that the use of any of these translational tools/methods has an impact on the governability of algorithmic systems. As such, we cannot yet know whether they help disadvantaged groups in society be heard and enabled to embed and protect their values in design tools, and then into the resultant AI systems. Consequently, we concluded that the existing translational tools and methods fail to operationalise AI ethics effectively. Almost all translational tools are either too flexible or too strict in the following sense (Arvan, 2018). When something (ethical tools, methods or guidelines) is too flexible it does little to protect against the risks of ethics shopping and ethics washing (Floridi, 2019b). In contrast, if the same something is too strict,  and approaches ethical governance in a top-down way, it fails to account for the fact that sometimes there is no social consensus about what is the 'right' way to interpret or apply ethics or ethical principles – this instead depends on how aggregate views of society are collected and which voices are included (Allen et al., 2000; Baum, 2017).  This overall conclusion (too flexible or strict) forces the AI ethics community to face the difficult question: if, even with technical guidance (such as that provided in IEEE's Ethically Aligned Design standards (IEEE Standards Association, 2019)) AI ethics is challenging to embed in the process of algorithmic Design, is the entire pro-ethical design (Floridi, 2019a) endeavour rendered futile? And, if no, then how can AI ethics be made useful for AI practitioners?

---

[1] Differentiating between public policy and research domains is important here. AI researchers have long been aware of the ethical implications of algorithms. Both Alan Turing and Norbert Wiener were writing on the topic as early as 1940. It has taken a longer time for policymakers, regulators and legislators to become interested in the topic.

In the following pages, we seek to answer these questions by exploring why principles and technical translational tools are still needed even if they are limited, and how these limitations can be potentially overcome by providing theoretical grounding of a concept that has been termed 'Ethics as a Service'[2]. Specifically, the sections 'lowering the level of abstraction' and  'limits of principlism and translational tools' explain in more detail the limitations of principlism and existing translational tools and methods. The section titled 'a series of compromises' outlines the compromises that must be made to enable the practical operationalisation of AI ethics. The section 'Outlining Ethics as a Service' provides the theory underpinning the concept of 'Ethics as a Service'. The final section concludes the article, highlighting where further research is needed.

---

[2] As will become clear through the development of the 'Ethics as a Service' concept in the following pages – our use of the concept is one grounded in the theory of Habermas's discourse ethics (Heath, 2014; Mingers & Walsham, 2010; Rehg, 2015) and Floridi's distributed responsibility (Floridi, 2016). This makes our interpretation of the concept distinct from the technocratic interpretation espoused by Google and other large tech firms claiming that they can 'audit customers' AI systems for ethical integrity' (Simonite, 2020). This paper should not, therefore, be read as being in support of such claims.

## 2. Lowering the level of abstraction

AI ethical guidance documents have been produced by a range of stakeholders, from technology companies, professional bodies and standards-setting bodies to governments and research organisations (Whittlestone et al., 2019). According to the Global Inventory of AI Ethics Guidelines, managed by Algorithm Watch, there are now more than 160 documents in existence (Alglorithm Watch, 2020). Whilst it is possible to summarise the principles contained within these documents as beneficence, non-maleficence, autonomy, justice and explicability (Floridi & Cowls, 2019), the range of concepts covered is vast and includes transparency; fairness; responsibility; privacy; freedom; trust; sustainability; dignity and solidarity (Jobin et al., 2019).

This variation, and consequential confusion, is perhaps to be expected. Many of the ethical harms that the principles in these documents purport to protect against are poorly understood because they are described too vaguely (Clarke, 2019). The vagueness of statements such as 'AI systems may be discriminatory' results in broad and generic rather than deep and specific responses. Additionally, as Carrillo (2020, p. 3) explains: 'beyond the basic underlying principles and common elements, ethical conceptions and principles vary across traditions, cultures, ideologies, systems and countries. In the end, if the expression 'ethics' in itself is universal, the content of 'the ethical' evolves and includes variable and flexible standards in accordance with the evolution of times and societies.' The risks that arise from this lack of clear ethical guidance are many and include: ethics washing; ethics shopping; ethics dumping; ethics shirking and ethics lobbying (Floridi, 2019b). Hence, ethical principles have been accused of being too flexible (or too undefined) to be of practical use to AI practitioners (Mittelstadt, 2019; Whittlestone et al., 2019). The accusation is mistaken insofar as the ethical principles should be seen as providing the foundation and not the details of ethical practices, in a way comparable to what a Constitution does when compared to specific legislation. It would be mistaken to criticise the Constitution of a country for being of no direct practical use in the regulation of medical appliances, for example. This is why a promising and reasonable approach to the problem of not-yet actionable AI ethical principles is to bring ethical guidance down to the Design level, by providing tools and methods that translate the 'what' of AI ethics into the 'how' of technical specifications. In doing so we can hope to create a bridge between abstract principles and technical implementations (Hagendorff, 2020). This is the solution that we explored in our previous research (Morley et al., 2019) and it is also the solution Digital Catapult are exploring in practice with the Digital Catapult AI Ethics Framework[3] (Box 1). In both this theoretical and applied work, we have concluded that this lowering of abstraction is, at best, a partial solution. Whilst translational tools and methods do help to lower the level of abstraction, they leave a number of other issues unresolved, and can be manipulated by reprehensible actors (Aïvodji et al., 2019).

In the following section we explore the limitations of translational tools in more detail. The Digital Catapult AI Ethics Framework (Box 1) can be considered an illustrative example of what we mean by 'translational tool'. The discussion is deliberately generalised and we recognise that some of the limitations we discuss can be overcome by combining the use of translational tools with other offerings. For example, the Digital Catapult's AI Ethics Framework is offered alongside more hands-on ethics consultations. This should be kept in mind so that we do not appear too critical and so that the motivation for us expanding on the concept of 'Ethics as a Service' in section 5 is clear.

## 3. Limits of Principlism and Translational Tools

The first limitation to highlight, is that translational tools and methods are extra-empirical. This means, as explained by Fazelpour and Lipton (2020), that they may set standards against which algorithmic practices are assessed, without themselves being subject to empirical evaluation. This leaves the translational tools vulnerable to manipulation. AI practitioners may choose the translational tool that aligns with what is for them the most convenient epistemological understanding of an ethical principle, rather than the one that aligns with society's preferred understanding (Krishnan, 2019). For example, certain types of 'explanation' can be used to obfuscate rather than illuminate (Aïvodji et al., 2019) patterns of injustice. In short, stated motivations for using a specific translational tool might not reflect actual motivation (Schiff et al., 2020).

---

[3]LF is chair of Digital Catapult's Independent Ethics Board. LK, EA and FG were employees of Digital Catapult at the time of writing. JM's work on applied AI ethics is partially funded by Digital Catapult.

Second, many of the existing translational tools and methods are 'diagnostic' but not 'prescriptive'. For example, they might identify whether a dataset is biased, but offer very little support to AI practitioners on how to overcome the issue. Others, as McMillan and Brown (2019) indicate, present 'technical fixes' to what are in essence socially-derived harms. Furthermore, when the parameters for the 'diagnosis' of fairness, transparency or accountability are set by the AI practitioners themselves, the potential for objective critique is easily lost and so the aim of the translational tool or method ceases to 'ensure the algorithmic system meets the criteria of social preferability' (Floridi & Taddeo, 2016) (and therefore ethical justifiability). Instead, the aim of the translational tool or method becomes to 'ensure the algorithmic system meets the practitioner's optimal criteria' (Terzis, 2020). As Martin (2019, p. 842) attests 'delegating a task to a technology [in this instance a 'translational' tool or method] does not remove the associated responsibility for that task. It is [still] a value-laden decision…' In short, according to Fazelpour and Lipton, (2020, p. 58), when used in this way, translational tools and methods: (a) can lead to systematic neglect of some [unethical] injustices and distort our understanding of others; (b) do not by themselves offer sufficient practical guidance about what should be done, sometimes leading to misguided mitigation strategies; (c) do not, by themselves, make clear who, among decision-makers is responsible for intervening to right specific [unethical] injustices (our additions in brackets).

Finally, too often these translational tools are positioned or perceived by AI practitioners as a 'one-off' test: something that just needs to be completed for compliance purposes (to be awarded a 'kitemark' of some description, for example) and then forgotten about. This encourages ethics by 'tick-box' and introduces the risk of writing ethics into the business case and coding them out by the time the algorithmic system is deployed (Morley et al., 2019). Instead, the ethical implications of an algorithmic system should be regularly evaluated, at a minimum as part of three distinct phases: validation, verification and evaluation. The first phase (validation) is concerned with whether the right algorithmic system is being developed; the second phase (verification) is concerned with whether the algorithmic system is being developed in the right way; and the third phase (evaluation) is concerned with whether the algorithmic system is continuing to operate in the right way once deployed, needs to be revised, or can be improved (Floridi, 2019a). Thus, unless ethical evaluation becomes an integral part of a system's operation (Arnold & Scheutz, 2018), there is no guarantee that pro-ethical translational tools will have any positive impact on the ethical implications of AI systems. Indeed, they could have a negative impact by fostering a false sense of security and consequential complacency.

This critique of AI ethics principlism and translational tools and methods raises the question whether the entire pro-ethical design endeavour is futile, if even with technical guidance, AI ethics is difficult or impossible to embed in the process of algorithmic design, development, deployment, and use. However, the experience of other applied ethics fields (for example, medical ethics and research ethics) shows that it is possible to operationalise abstract ethical principles successfully for the purpose of protecting individuals, groups, society and the environment from particular social harms and incentivising the best outcomes. The effort is not futile. With this reassurance in mind, the next pertinent question becomes: how can AI ethics be usefully operationalised for AI practitioners? The next section offers a way forward.

**Box 1:** *The Digital Catapult AI Ethics Framework (DCEF)*

The DCEF was developed by the Digital Catapult's independent ethics board following consultation with a number of Digital Ethicists and other experts. The framework consists of four levels, and is intended to help AI start-ups working with the Digital Catapult to define and translate, transparently and contextually, high-level ethical principles into practice. The first level, therefore, consists of the five unifying high-level principles identified by Floridi et al., (2018): beneficence, non-maleficence, autonomy, justice, explicability. The second level consists of seven interpretations (or contextual definitions) of these principles identified through documentary analysis consultation with AI practitioners and those affected by AI systems. The third level operationalises Habermas's concept of discourse ethics (Buhmann et al., 2019), i.e. an approach that seeks to establish normative values and ethical truths through open discourse, and consists of a series of questions that are designed to encourage AI practitioners to conduct ethical foresight analysis (Floridi & Strait, 2020). The fourth level provides access to more practical, and less discursive tools e.g. python libraries designed to identify bias in data. The connections between the levels are shown below. Companies using the DCEF to translate high-level ethical principles into practice are encouraged to consult it at validation, verification and evaluation stages of their product development pipeline, to ensure that at each stage time is dedicated to thinking through the ethical implications of all decisions made. This discussion is supported by members of the independent[4] ethics board through consultations which also provide a vehicle for reviewing the efficacy of the Framework itself.

| L1 | **Beneficence:** promoting well-being, preserving dignity, and sustaining the planet. | **Non-maleficence:** privacy, security and 'capability caution.' | **Autonomy:** the power to decide (whether to decide). | **Justice:** promoting prosperity and preserving solidarity. | **Explicability:** enabling the other principles through intelligibility and accountability. |
|---|---|---|---|---|---|
| L2 | Be clear about the benefits of the product or service.<br><br>Consider the business model. | Know and manage the risks.<br><br>Use data responsibly | Be open and understandable in communications. | Promote diversity, equality and inclusion. | Be worthy of trust. |
| L3 | For example:<br><br>What are the goals, purposes and intended applications of the product or service?<br><br>Who or what might benefit from the product/service? Consider | For example<br><br>Is the training data appropriate for the intended use?<br><br>Have potential biases in the data been examined, well-understood and documented and is there a | For example:<br><br>Does the company communicate clearly, honestly and directly about any potential risks of the product or service being provided? | For example:<br><br>Are there processes in place to establish whether the product or service might have a negative impact on the rights and liberties of individuals or groups? | For example:<br><br>Is there a process to review and assure the integrity of the AI system over time and take remedial action if it is not operating as intended? |

---

[4] By 'Independent' we mean board where none of the members are employees of the AI company in question but are still embedded within the company to a sufficient degree as to be able to have access to necessary documentation, data and code, and understanding of the socio-technical context ('International AI Ethics Panel Must Be Independent', 2019; Raji et al., 2020). If necessary, this can be managed by placing the board members under non-disclosure agreements.

| | | | | | |
|---|---|---|---|---|---|
| | all potential groups of beneficiaries, whether individual users, groups or society and environment as a whole. | plan to mitigate against them? | Are the company's policies relating to ethical principles available publicly and to employees? Are the processes to implement and update the policies open and transparent? | Does the company have a diversity and inclusiveness policy in relation to recruitment and retention of staff? | Does the company have a clear and easy to use system for third party/user or stakeholder concerns to be raised and handled? |
| L4 | See: https://www.digicatapult.org.uk/for-startups/other-programmes/applied-ai-ethics-typology | | | | |

Box 1: The Digital Catapult AI Ethics Framework. The full framework is available here:
https://www.digicatapult.org.uk/for-startups/other-programmes/ai-ethics-framework

## 4. A series of compromises

Thus far we have seen that the need to Design AI solutions pro-ethically is well recognised, and that the field of AI-ethics now has a solid foundation comprised of principle-based governing documents and translational tools and methods. Developing this foundation has been essential and the individual components remain highly valuable. However, pro-ethical Design practices remain difficult to operationalise practically as the Goldilocks Level of Abstraction has not yet been found. Attempts thus far remain either too flexible or too strict. To overcome these limitations, the 'just right' Level of Abstraction needs to be identified by making a series of compromises.

### 4.1 Finding a compromise between too flexible and too strict

Raab (2020) argues that top-down, prescriptive guidelines imply that it is possible to take a formulaic approach to the application of ethical norms, principles and general rules to specific instances. In reality, the argument continues, applied ethics requires judgement. Specifically, it requires an ability to consider how risks, conflicting rights and interests, and social preferability varies depending on a particular context. The ethical implications of deploying an AI system in a healthcare setting are unlikely to be the same as the ethical implications of deploying an AI system in an educational setting. Similarly, the boundaries of social preferability within Europe may not be the same as the boundaries within Asia and these boundaries may change with time or with type of algorithm, or even with stage of development. Finally, ethical guidelines that are too strict portray algorithmic systems as static products of code and data that once deployed continue to operate in the same way as intended and have only the intended (positive) effects. Ananny & Crawford, (2018) point out the reality is that algorithmic systems are assemblages of human and non-human actors which have many non-deterministic impacts. To understand (and therefore govern) the ethical implications requires understanding how the whole system works – including what may happen once the system is deployed and used by entities, or for purposes, other than the original practitioners or stated purpose.

With this in mind, if AI ethics is to be operationalised in a way that is useful to AI practitioners and simultaneously protective of individuals, groups, society and the environment, then the operationalisation must: (a) happen at the appropriate Level of Abstraction (where translational tools are neither too flexible, nor too strict); and (b) must not consist solely of a one-off tick-box exercise completed only at the beginning of the Design process. Developing a practical pro-ethical Design approach that meets these two criteria is not simple, but it is not impossible. It requires a shift in the way that AI ethics is framed. The practice of AI ethics should not be seen as an end-goal that can be objectively achieved, observed, quantified, or compared. Instead it should be seen as a reflective development process, which also aims to help AI practitioners understand their own subjectivity and biases within a given set of circumstances (Terzis, 2020). By enabling the development of this understanding, a reflective process can help illuminate why unethical outcomes may occur so that the appropriate mitigation or avoidance strategy can be put in place (Fazelpour & Lipton, 2020). From this perspective, the practical operationalisation of AI ethics becomes less about the paternalistic imposition of inflexible standards that ignore context and more about procedural regularity and public reason that can be adapted and shared across contexts and societies (Binns, 2018; Kroll et al., 2017). In practice, structured identification and transparent communication of tradeoffs help organisations arrive at resolutions that, even when imperfect, are at least publicly defensible (Whittlestone et al., 2019).

An operationalisation of AI ethics focused on procedural regularity and public reason would commit a company producing algorithmic systems to:

a) justifying all design decisions to a set of common principles agreed through an inclusive and discursive process that involves all individuals, groups and environmental representatives likely to be directly or indirectly affected by the AI products of a specific company or research group. These principles should be reviewed periodically (e.g., annually);

b) following a set and repeatable procedure to define and translate each of the agreed upon principles into technical standards in a way that achieves an acceptable level of ethical justifiability and environmental sustainability within the specific context; and

c) ensuring appropriate oversight is in place at the validation, verification and evaluation stages.

Companies operationalising AI ethics in this way would use translational tools and ethical principles in the same way each time (and at repeated intervals to cover validation, verification, evaluation) but accept that the exact way in which they are applied is contextually dependent. This kind of shift away from abstract rules towards negotiated ethics has already been seen in arguments for a move from privacy by design to privacy engineering. Privacy engineering, as described by Alshammari & Simpson (2017, p. 162), is a 'means of applying engineering principles and process in developing and maintaining systems in a systematic and repeatable way, with a view to achieving acceptable levels of privacy protection' without assuming that the way that this is achieved will be the same in each instance or immutable through time or different contexts. Additionally, this approach to operationalisation covers the five pillars of good ethical governance set out by Winfield & Jirotka (2018) by turning AI ethics into a 'collaborative process, developed and iteratively (re)configured through material practices and continued negotiations' (Orr & Davis, 2020, p. 731). However, if the responsibility for the whole process still sits with just the AI practitioners themselves, there remains a risk that the operationalisation process itself becomes subject to manipulation and may be used solely for ethics washing purposes.

## 4.2 Finding a compromise between devolved and centralised responsibility

One often highlighted option for avoiding these potential issues – derived from a lack of accountability and transparency – is to rely on external algorithmic audits (Holstein et al., 2018; Mökander et al., forthcoming). In theory, the process described above could be managed internally by the company in question but audited by a third party (Mökander et al., forthcoming). Several auditing mechanisms have been proposed as means of examining the inputs and outputs of algorithms for bias and other harms (Cath, 2018; Sandvig, 2014). For example, 'Aequitas' is an open source toolkit which audits algorithms for bias and fairness (Saleiro et al., 2018) and 'Turingbox' is a proposed platform that would audit the explainability of an algorithmic system (Epstein et al., 2018). As these technical methods, and more human-based methods such as 'sock-puppet' auditing (Sandvig, 2014), have gained visibility, the perceived importance of the role they will play in ethical governance has also increased. In the UK, for example, the Information Commissioner's Office is developing an AI auditing framework that seeks to ensure organisations have measures in place to be compliant with data protection requirements, and mitigate risks associated with (amongst other issues) fairness, accuracy, security, and fundamental rights (Binns, 2018). Similarly, auditing firm PwC includes ethical audit as a key component of its responsible AI Framework (Oxborough et al., 2019). It is clear, therefore, that external ethical auditing will be a key component of any form of operationalised AI ethics.

However, there remain limitations. Proposed external auditing mechanisms typically focus on specific parts of the system, rather than looking at the overall system function (Cath, 2018), and so do little to address the risks of 'unpredictability' described above. Audits cannot guarantee to reveal all aspects of a system (Kroll, 2018), and so may miss the inputs that are the most harmful. In addition, they are typically conducted after a system has already been deployed, and so may have already had a negative impact (Raji et al., 2020). And, finally, audits may be rendered not viable without legislative change due to legal concerns regarding protection of consumer data or trade secrets (Katyal, 2019; Kroll et al., 2017). Moreover, it is difficult to quantify indirect externalities that accumulate over time (Rahwan, 2018). External auditors may therefore not only lack access but also resources, know-how and computational power to review AI systems (Kroll, 2018). Raji and colleagues (2020) discuss these limitations in detail and also note that the agile nature of AI development and typical lack of documentation challenges auditability. Furthermore, they stress how the lack of foresight analysis typically included in external post-hoc audits minimises the chances for audits to prevent future harms. They argue that internal audits, conducted by a dedicated team of organisational employees – but not the AI practitioners themselves – with full access to data and a focus on ethical foresight could be a pragmatic alternative. There is certainly some promise in this proposal. Code review and internal checks for reliability and robustness are already common practices within software engineering. Furthermore, internal auditing of this nature would mitigate the risks associated with external auditing (or fully-external ethical review boards) of ethically desensitising, de-skilling, and de-responsabilising company employees, and instead force companies to make their own critical choices and assume explicit responsibilities (Floridi, 2016). Yet, it is also undeniable that internal auditors may face conflicts of interest that make it hard for them to maintain an independent and objective opinion (Raji et al., 2020). This is especially true when there are currently limited incentives for companies to rigorously examine the implications behind technologies that are both profitable and powerful (Katyal, 2019). Thus, just as it was necessary to find a compromise between mechanisms that are too fleixible or too strict, it seems that there is also a need to find a compromise between completely devolved and completely centralised responsibility for holding the ethical governance process itself accountable.

The solution here lies in the creation of a multi-agent system where the responsibility is distributed across different agents (individuals, companies) in a way that (a) aggregates the possibly good actions, so that the latter might reach the critical mass necessary to make a positive difference to the targeted environment and its inhabitants, and (b) isolates possibly negative actions (e.g., attempts to ethics wash), so that they never reach the necessary threshold to breach the fault-tolerance level of the overall system and undermine its effectiveness as an ethical governance mechanism (Floridi, 2013, 2016). This may seem overly theoretical, but it is actually a practically feasible approach, and a pragmatic analogy can be found in cloud computing, as explained in the next section.

## 5. Outlining Ethics as a Service

Cloud computing, the on-demand delivery of various computing services over the internet, has three models of service: Software as a Service (SaaS), Infrastructure as a Service (IaaS), and Platform as a Service (PaaS).

Software as a Service is the model of cloud computing that consumers are most readily familiar with. All aspects of the service are managed by a third-party, there are few customisation opportunities, and often there is a high risk of vendor lock-in. In our analogy regarding AI ethical governance, this would represent the fully devolved model of governance (and one that is too strict). A third party would be responsible for dictating the set of ethical principles, for outlining the process that must be followed at each of the validation, verification, evaluation stages, and for conducting an ethical audit to see whether the process was followed correctly and whether this resulted in the expected positive outcomes.

Infrastructure as a Service, in comparison, represents the fully centralised governance model (and one that is too flexible). In cloud computing, servers, network operating systems, and storage are all provided via a dashboard or application programme interface (API), so that users have complete control over the entire infrastructure. In terms of ethical governance, this would involve the AI practitioners being responsible for both developing the AI ethics principles as well as the process to follow internally with limited meaningful engagement with external stakeholders. The AO practitioners would also be responsible for conducting internal audits.

Finally, there is Platform as a Service which represents the compromises we have outlined above. It is the Goldilocks option found between methods that are too flexible and strict, and between devolved governance and centralised governance. In the world of cloud computing, PaaS represents a set-up where the cloud provider provides the core infrastructure, such as operating systems and storage, but users have access to a platform that enables them to develop custom software or applications. These three options are summarised in Figure 1.

|  | Centralised Responsibility (too flexible) (Infrastructure as a Service) | Distributed Responsibility (Ethics as a Service) (Platform as a Service) | Devolved Responsibility (too strict) (Software as a Service) |
| --- | --- | --- | --- |
| Internal Responsibility | Developing ethical Code and contextually defining meaning of each principle<br>Evaluating selecting and using translational tools and methods<br><br>Conducting ethical review of own product at 3 stages and including ethical foresight analysis<br>Auditing AI systems for ethical compliance and social impact | Contextually defining meaning of each principle within the ethical Code<br>Selecting and using translational tools/methods from a pre-approved list of available translational tools/methods<br>Conducting ethical review of own product at 3 stages and including ethical foresight analysis | Using the translational tools/methods they are told to |
| External Responsibility | N/A | Developing ethical Code, regularly reviewing it and developing a set process that AI practitioners must follow to contextually define principles<br>Evaluating available translational tools/methods and compiling a pre-approved list available for selection and use by the internal AI practitioners and developing a process that AI practitioners must follow to contextually to select which of the pre-approved translational tools/methods they will use<br>Auditing AI systems for ethical compliance and social impact | Developing ethical Code and contextually defining the meaning of each principle included in the code.<br><br>Evaluating and dictating exactly which translational tools and methods can be used by the internal AI practitioners<br><br>Conducting ethical review of product at 3 stages and including ethical foresight analysis as part of overall auditing of AI systems for ethical compliance and social impact |

Figure 1: Comparison of distributions of responsibility for ethics-related activities in different AI ethics governance models. Centralised responsibility and devolved responsibility models represent the status quo, the Ethics as a Service model is the new proposal.

In the world of AI ethical governance, Ethics as a Service – based on Platform as a Service model – could involve several components including, but not necessarily limited to: an independent multi-disciplinary ethics board; a collaboratively developed ethical code; and AI practitioners themselves. Responsibility could then be distributed across these components thus:

1. **Independent multi-disciplinary ethics advisory board** responsible for providing the core infraethics as described:

a. The development of a principle-based ethical code through a process of discussion and negotiation that treats ethical patients (i.e. individuals, businesses and environments that may be impacted by the systems produced by the AI company in question) as real interlocutors, who can genuinely impact the design of the system(Aitken et al., 2019; Arvan, 2018; Durante, 2015).

b. The setting out of a process that needs to be followed at validation, verification and evaluation stages of algorithmic Design to ensure pro-ethical design by: (i) defining contextually the specific meaning of each of the principles in the ethical code; and (ii) providing an appropriate range of proven-effective translational tools which can be selected to translate from principles to practice, according to the contextually specific definitions. This process and selection of translational tools must include an element of ethical foresight analysis (Floridi & Strait, 2020), and mechanisms for closing down, and rectifying the consequences of a system that is found to be in breach of the principles (Morley et al., 2020). It must also recognise that positive, ethical features are open to progressive increase, that is an algorithm can be increasingly fair, and fairer than another algorithm or a previous version, but makes no sense to say that it is fair or unfair in absolute terms (compare this to the case of speed: it makes sense to say that an object is moving quickly, or that it is fast or faster than another, but not that it is fast in absolute terms).

c. Conducting regular audits of the whole behaviour of the company – not just the end product once launched – to see whether it is genuinely committed to ethical conduct; whether AI practitioners are following the defined process; and whether the final output is ethically justifiable according to contextually-defined principles.

2. **The internal company employees** (the AI practitioners themselves) responsible for providing the 'customised software,' namely:

   a. Contextually defining the principles;

   b. Identifying the appropriate tools, and putting them to use whilst designing a specific algorithmic system;

   c. Documenting how the process was followed, in public, and justifying why specific decisions were made, when and by who.

In theory, distributing the responsibility for operationalising AI ethics in this manner will overcome many (although definitely not all) of the limitations of current approaches described above. However, whether it works in practice is yet to be seen. It is also important to note that, because AI systems learn and update their internal decision-making logic over time, code-audits (as described in 1c above) need to be complemented with continuous functionality and impact audits. However, such audits inevitably impose both financial and administrative costs (Brundage et al., 2020). Care should therefore be taken to not put undue burden on certain sectors in society (Koene et al., 2019). One way to balance the need for audits with incentives for innovation is to introduce a progressive level of AI governance that is proportional to the risk level associated with a specific combination of technology and context. Therefore, testing of the concept must involve experimentation to find the proportionate degree of oversight for different AI solutions. For this reason (and others), further research is urgently needed on how to evaluate translational tools and, in doing so, evaluate the current impact of the AI ethics endeavour in order to highlight further ways in which it could be improved. To start this research, a partial pilot of "Ethics as a Service," which includes the Digital Catapult AI Ethics framework (Box 1), ethics consultations and an Independent multi-disciplinary ethics advisory board is being trialed by Digital Catapult. We would encourage others to develop partial or complete pilots of the concept and to publicly report on the successes and failures so that a commons of knowledge related to 'best ethical practice' can be established.

## 6. Conclusion

As Thomsen (2019) states, 'ethics for AI cannot be expected to be any simpler than ethics for humans.' Indeed, it may be more complicated, since it adds to it further technical issues. Research ethics and medical ethics have always involved a combination of the law, ethical governance policies, practices, and procedures, with contextual discursive and procedural support. This combination approach has enabled these branches of applied ethics to find a good balance between being too strict and too flexible , and between too centralised and too devolved. Therefore, it seems reasonable to hypothesise that AI ethics would benefit from an equally customisable approach, and that if this balance can be achieved then the pro-ethical Design endeavour may succeed. At the very least shifting the focus of AI ethics away from principles to procedural regularity will make AI ethics seem more relatable to AI practitioners. Encouraging a procedural approach can, for example, help make the parallels between AI ethics and other quality assurance processes, such as safety testing, clearer and thus make it more obvious why careful consideration needs to be given to each Design decision. We hope that the idea of Ethics as a Service, as outlined in the article, has at least highlighted this.

Whilst these opportunities for moving forward the conversation about AI ethics and the role that Ethics as a Service may play in this, should be celebrated, it must be acknowledged that the impacts of AI systems cannot be entirely controlled through technical design (Orr & Davis, 2020). Biased AI is not simply the result of biased datasets, for example. AI solutions themselves are complex and are then deployed into complex systems. In complex systems, agents interact with each other and with other systems in unexpected ways, making their response to change unpredictable and non-linear. It is, therefore, likely that we will genuinely not know whether any approaches to 'pro-ethical' Design have made an impact (positive or negative) on the social impact of an algorithmic system until after it has been deployed. Regular re-evaluation of all aspects of algorithm systems, and the extent to which they achieve their goals, including pro-ethical Design approaches, will be crucial. Hence, further qualitative research and empirical testing will be needed to understand in detail the benefits, and drawbacks, as well as the practicalities of Ethics as a Service. This will be our next task.